# Room-Temperature Deuterium Separation in van der Waals Gap Engineered Vermiculite Quantum Sieves

*Saini Lalita[1], Rathi Aparna[1], Kaushik Suvigya[1], Li-Hsien Yeh[3,4] and Kalon Gopinadhan[1,2]\**

[1]*Department of Physics, Indian Institute of Technology Gandhinagar, Gujarat 382355, India*
[2]*Department of Materials Engineering, Indian Institute of Technology Gandhinagar, Gujarat 382355, India*
[3]*Department of Chemical Engineering, National Taiwan University of Science and Technology, Taipei 10607, Taiwan*
[4]*Advanced Manufacturing Research Center, National Taiwan University of Science and Technology, Taipei 10607, Taiwan*

*(\*Correspondence: gopinadhan.kalon@iitgn.ac.in)*

**Abstract:**

As the demand for nuclear energy grows, enriching deuterium from hydrogen mixtures has become more important. However, traditional methods are either very energy-intensive because they require extremely cold temperatures, or they don't separate deuterium ($D_2$) from regular hydrogen ($H_2$) very well, with a $D_2/H_2$ selectivity of about ∼0.71. To achieve efficient deuterium separation at room temperature, we need materials with very tiny spaces, on an atomic scale. For the first time, we've successfully created a material with spaces just about ∼2.1 Å (angstroms) wide, which is similar in size to the wavelength of hydrogen isotopes at room temperature. This allows for efficient deuterium separation, with a much higher $D_2/H_2$ selectivity of ∼2.20, meaning the material can separate deuterium from hydrogen much more effectively at room temperature. The smaller deuterium molecules are more likely to pass through these tiny spaces, showing that quantum effects play a key role in this process. In contrast, a material like graphene oxide, with larger spaces (around 4.0 Å), only shows a lower $D_2/H_2$ selectivity of ∼1.17, indicating weaker quantum effects. This discovery suggests that materials with very small, atomic-scale spaces could be key to efficient separation of hydrogen isotopes at room temperature.

1. Introduction

One of the most promising avenues of renewable energy, nuclear reactors require deuterium in heavy water as a moderator or fuel for potential fusion reactors[1]. However, deuterium is scarce, and its extraction has always been challenging. Conventionally, hydrogen isotope separation is done by processes like Girdler-sulfide and cryogenic distillation[2], and they show very low $D_2/H_2$ selectivity. Given the use of deuterium in other applications[3] like medicine (as tracers) and spectroscopy (for contrast enhancement), the demand for deuterium and other deuterated products is expected to go up. For such applications, deuterium is of greater interest than hydrogen, and hence, separation techniques that permeate deuterium more efficiently than hydrogen are ideal. A selectivity $D_2/H_2 > 1$ means the selectivity toward deuterium, which is not the case for the conventional methods as they are all selective to hydrogen and hence have a $D_2/H_2 < 1$.

Recent efforts to find efficient hydrogen isotope separation mechanisms geared towards quantum tunneling and quantum sieving[4]. These mechanisms apply to particles under confinement but affect the isotope separation in strikingly different ways. Briefly, under confinement, when particles encounter a thin energy barrier greater than their kinetic energy, they can still cross the barrier by quantum tunneling, with the lighter hydrogen, $H_2$, being more efficiently transported than the heavier deuterium, $D_2$. Here, the transport efficiency, $T_e \propto \exp(-\sqrt{2M\phi}d/\hbar)$, where $M$ is the isotope mass, $\phi$, the barrier energy, $d$ the barrier thickness and $\hbar$ the reduced Plank's constant. The above expression implies that ultrathin membranes are essential for achieving significant separation based on quantum tunneling.

On the other hand, when the confinement is extreme and comparable to the de Broglie wavelength ($\lambda = h/\sqrt{3Mk_BT}$, where $T$ is the temperature and $h$ is the Plank's constant) of the particles, the nuclear quantum effects become more dominant[4]. The nuclear quantum effects are predicted to affect the adsorption (chemical affinity quantum sieving) and transport (kinetic quantum sieving) of isotopes under such confinements. In this case, the confinement leads to a different kind of isotope separation, where the heavier isotope is transported/ adsorbed preferentially due to its smaller wavelength. In such cases, the de Broglie wavelength associated with the particle becomes the differentiating factor of the isotopes rather than their mass. For the hydrogen isotopes $H_2$ and $D_2$, at a temperature of 30 K, $\lambda_H$ and $\lambda_D$ is 4.6 Å and 3.2 Å, respectively. The small size implies that angstrom-sized transport channels are necessary for observing quantum sieving effects. Such sizes were realized in porous materials like MOF[5–8], Zeolites[9–11], $Ti_2O_3$[12], etc., and led to the observation of chemical affinity quantum sieving via selective adsorption of isotopes[4]. While promising in terms of selectivity, these are still energy-intensive due to the requirement of cryogenic temperatures.

Realizing quantum sieving at room temperature (300 K) requires smaller fluidic channels, as $\lambda_H$ and $\lambda_D$ are 1.5 and 1.0 Å, respectively. The naturally occurring and externally controllable angstrom-sized pores and interlayer spaces in van der Waals materials are favorable for observing quantum effects at moderate temperatures. Recently, there have been reports on controllable proton transport through different 2D materials by methods like gate[13,14] and corrugations[15], which helped in understanding the proton transport through confined spaces of 2D materials. In the realm of hydrogen isotope separation, it is equally important to understand the nature of deuterium transport as the difference in their de Broglie wavelengths translates to different adsorption or transport rates.

On this front, very few studies report transport across interlayer spaces of single crystals of hBN, $MoS_2$[16] and defects (pores) in pyridinic N-doped graphene[17]. The $D_2/H_2$ selectivity in these reports is close to unity. Whereas these reports are optimistic for room-temperature processes, the selectivity still needs to be improved, and the sample fabrication processes are tedious and expensive. Therefore, exploring novel materials with better selectivity and cost-effective fabrication technologies is highly required.

Among the diverse properties of 2D materials, the intercalants inserted in interlayer spaces limit the spaces available to transport particles. They can help achieve the required confinement for observing room temperature quantum sieving effects. Herein, we report a large room-temperature $D_2/H_2$ separation based on kinetic quantum sieving through deuterium intercalated vermiculite laminates exhibiting a van der Waals gap of ~2.1 Å. Vermiculite is a layered clay material that can easily be exfoliated into sheets and assembled into laminates[18]. The interlayer spaces of these laminates can be controlled by cation-intercalation.

Quantum sieving-based isotope separation with van der Waals crystals is rarely reported. As far as our knowledge is concerned, there has been no study with micron thick laminates made from exfoliated two-dimensional materials. We used two complementary techniques- electrical conductivity and mass spectrometry to confirm the observation of significant isotope selectivity.

2. **Results and Discussion:**

In our previous work[19], we controlled the effective interlayer spacing, also known as the van der Waals gap in vermiculite laminates in the range 3-5 Å by intercalating mono, bi, and tri-valent cations. However, a channel size of ~2 Å or lower is required to observe room temperature quantum sieving. To achieve this, we intercalated smaller ions $H^+$ and $D^+$ in the vermiculite interlayers using hydron (proton or deuteron) selective Nafion polymer and electric field. As detailed in the Supporting Information, a platinized carbon cloth acts as the hydron source electrode for facilitating the intercalation process **(Figure 1a)**. Briefly, for the intercalation of vermiculite with protons (deuterons), the devices were kept in a hydrated (deuterated) environment in the presence of $H_2$ ($D_2$) and a voltage of 0.1 V was applied across a laminate of thickness 3 μm **(Figure S1a, Supporting Information)** for 5 hours. The intercalation is confirmed using X-ray diffraction **(Figure 1b)**. For the proton-intercalated vermiculite (H-V-N), we observed an interlayer spacing of ~13.5 Å, very close to that of vermiculite intercalated with hydrochloric acid **(Figure S1b, Supporting Information)** or acetic acid[20]. However, the actual space available for the transport (known as the van der Waals (vdW) gap) is much smaller, estimated by subtracting the thickness of silicate layers (9.6 Å) [21]. The vdW gap of the H-V-N sample is thus estimated to be ~3.9 Å, which is pretty large enough to see any significant quantum sieving at room temperature.

To our surprise, the deuteron intercalated vermiculite (D-V-N) showed a smaller interlayer spacing of 11.7 Å, leaving a vdW gap of ~2.1 Å for the transport, adequate for observing quantum sieving effects **(Figure 1c)**. Interestingly, this effective interlayer spacing does not change back to that of H-V-N even after being exposed to hydrated environments again (**Figure S1c, Supporting Information**), which leaves the vdW gap fixed at ∼2.1 Å. This cation fixation has been observed in vermiculite with the isotopes of heavier

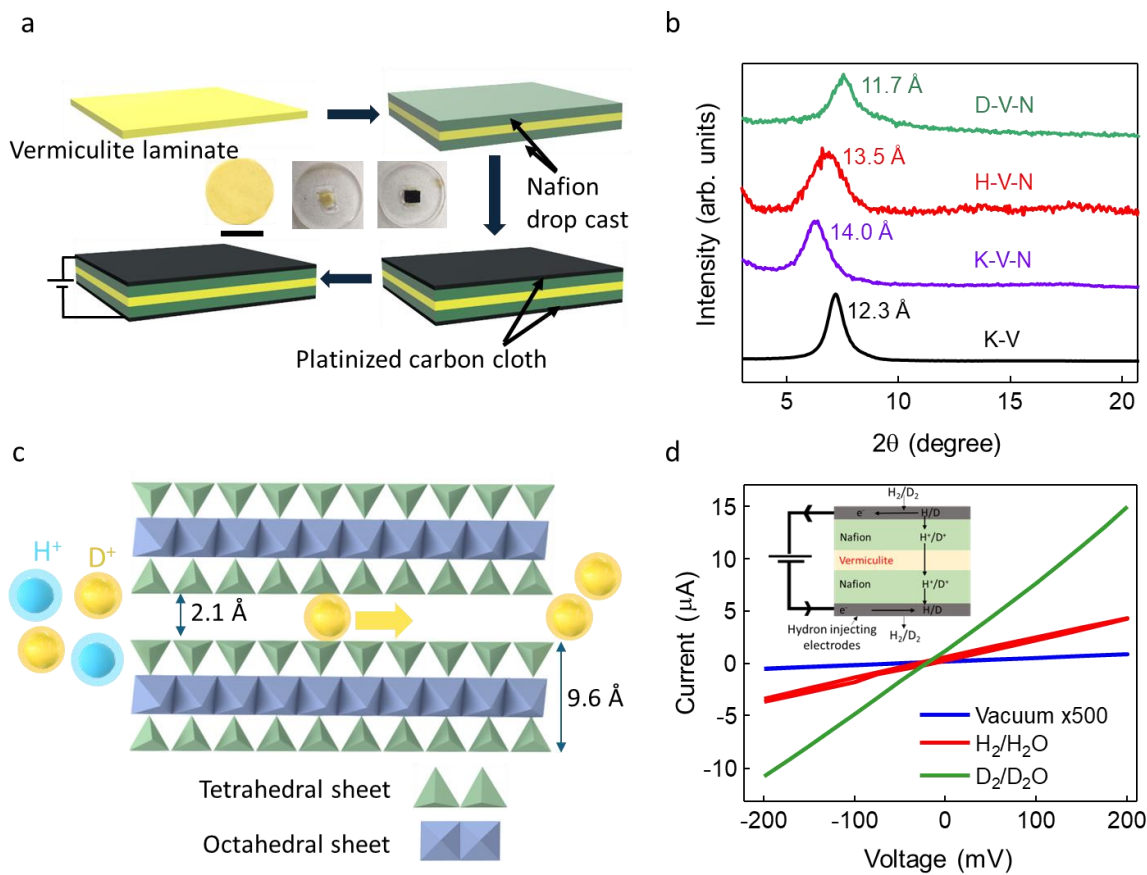

**Figure 1. Fabrication of vermiculite devices and their characterization.** (a) Schematic diagram of the vermiculite laminate device fabrication steps for proton/deuteron transport. We first coated the laminate with Nafion (proton conducting layer). Then, we attached the platinized carbon cloth on both sides as the electrodes (in the middle- camera images captured at various stages of sample preparation. Scale bar = 1 cm). (b) X-ray diffraction data of vermiculite laminates exchanged with potassium (K-V, black), which shifts toward lower angles when coated with Nafion (K-V-N, purple), hydrogen exchanged vermiculite coated with Nafion (H-V-N, red), and deuterium exchanged vermiculite coated with Nafion (D-V-N, green). The deuterium exchange decreased the interlayer spacing to 11.7 Å and fixed the spacing. (c) Schematic representation of hydrogen isotope transport through interlayer spaces of vermiculite laminate. The vermiculite sheets are made up of octahedral sheets (blue) sandwiched between two tetrahedral sheets (green). The vdW gap has been controlled to 2.1 Å by intercalation of deuteron. We represent the de Broglie wavelength in terms of size, with the deuteron (yellow) smaller than the proton (light blue). The yellow arrow indicates the easier transport of deuterium over the hydrogen (d) The current-voltage characteristics of vermiculite device for proton (deuteron) transport. As a reference, we also plotted

the data recorded under vacuum conditions. (Inset: schematic diagram of proton and deuteron transport through vermiculite devices).

elements like Cesium[22,23] but has never been reported for lighter elements like hydrogen. While the proton intercalation in clays is well known, little information is available on deuterium intercalation. We believe that the strong electrostatic interaction between cations- protons and deuterons and the negatively charged vermiculite layers stabilize the structure. The reduced interlayer space with deuteron intercalation and its preferential intercalation over proton can only be understood if we invoke the quantum nature of these particles. The smaller de Broglie wavelength of the deuteron enables its preferential adsorption over the proton. The deuteron has less tendency for deuteration (due to smaller nuclear potential) than proton hydration and hence the layers are electrostatically much more active, thus causing a shrinkage in the interlayer space. The preferential adsorption of deuterium over hydrogen has also been observed in other materials[6,11], which is explained based on the chemical affinity quantum sieving.

Having controlled the vdW gap to ~2.1 Å, we investigated the hydron transport through our devices (**supplementary section 1, Supporting Information**). For all the measurements, we placed the devices in an environment of $H_2$ (100%), $D_2$ (100%) or its mixture (**Figure S2, Supporting Information**). The humidity ($H_2O$ or $D_2O$) inside the chamber was maintained at 100% throughout the measurements to ensure the proper wetting of the Nafion. To ensure the accuracy of our experimental setup, we first measured the transport of protons and deuterons across reference samples consisting of commercial Nafion 117 only. We applied a voltage (*V*) across the sample, which resulted in a finite electrical current (*I*). The proton conductance, $G = I/V$, showed values similar to the previous report[24], and higher than the deuteron conductance (**Figure S3, Supporting Information**). Next, we measured the vermiculite samples in an $H_2/H_2O$ environment. As Nafion offers very high proton and negligible electron conductivity, the measured *G* directly corresponds to the proton permeation through our samples. The linear variation of current with voltage (**Figure 1d**) indicates the absence of capacitive interfaces in our devices. Taking the thickness of Nafion (180 µm) and vermiculite (3 µm) into account, the proton conductivity ($k = Gl/A$, where *l* is the thickness and A is the area) of the reference sample was about four orders higher than that of vermiculite samples (**Figure S3, Supporting Information**). This result implies a negligible contribution of Nafion coating on the conductance measured through vermiculite samples. We believe that the main component contributing to the resistance through vermiculite is the energy barrier at the entrance of the channels. Measuring *G* of vermiculite samples over a range of temperatures confirmed that such a barrier indeed exists, and the activation energy ($E_a$) can be calculated from the Arrhenius relation:

$$G \propto exp\left(-\frac{E_a}{k_B T}\right) \qquad \qquad \text{Equation (1)}$$

where $k_B$ and $T$ are the Boltzmann constant and the temperature, respectively.

After the $H_2/H_2O$ experiment through a device, we replaced the $H_2/H_2O$ environment with $D_2/D_2O$ to study deuteron transport through the same device. This method effectively replaces all the $H^+$ to $D^+$ and makes Nafion function as a deuteron-conducting polymer[16], which allows us to measure the activation energy for the deuteron barrier **(Figure 2a)**. Following the same process mentioned above, $E_a^D$, we estimated the activation energy for the deuteron barrier for vermiculite devices, which came out to be lower than that of protons. For quantifying the transport efficiency of proton versus deuteron, we calculated the $D_2/H_2$ selectivity, $\alpha$, with the help of the following relation:

$\alpha = G_D/G_H = \exp(\frac{\Delta E_a}{k_B T})$, where $\Delta E_a = E_a^H - E_a^D$ is the difference in the activation energy of proton and deuteron. Due to the transport's similar nature, the pre-exponential factor of the Arrhenius equation remains the same for both these isotopes, as confirmed by the mass spectrometer measurements described later in the text. We recorded a range of selectivity (4.68 to 1.78) for various vermiculite devices **(Figure 2b)** with an average of around 2.20 at room temperature. We cannot explain the observation of $D_2$ transporting more than $H_2$ by any mechanisms other than quantum sieving. Such a high $D_2/H_2$ selectivity value is remarkable and indicates the great potential of vermiculite laminates for isotope separations.

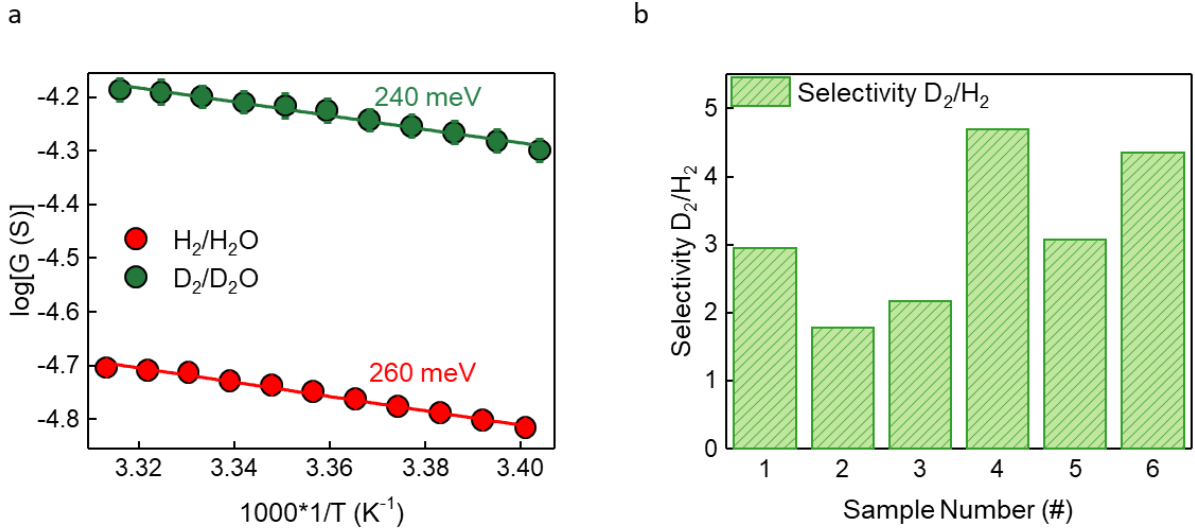

**Figure 2. Electrical detection of protons/deuterons through vermiculite devices**. (a) Estimation of activation energy of proton versus deuteron across a vermiculite device from the temperature (*T*) dependent conductance (*G*) measurement. The error bars represent the standard deviation of the data. (b) The values of $D_2/H_2$ selectivity for six different samples.

Many recent studies use mass spectrometer [25] to analyze the gas mixtures,[26] protons[27] and hydrogen isotopes[17,24] permeating through polymeric[25] or 2D membranes[28]. This technique provides real-time monitoring of the permeating species through current/partial pressure. By observing the signal of permeating species compared to the known feed mixture, one can easily calculate selectivity by using the current/partial pressure values as detected by the mass spectrometer.

We utilized a single quadrupole mass spectrometer (Cirrus 3XD, MKS) to analyze the gases transporting through the vermiculite interlayers to corroborate the findings of electrical measurements. We calibrated the mass spectrometer's sensitivity using known mixtures and monitored their ion fragments to measure the permeated species accurately[29,30]. More details regarding the calibration are given in the supporting information (supplementary section 2).

The device setup for the mass spectrometry is quite similar to the electrical conductance measurement, except that a mass spectrometer is used on the permeate side. More details are given in the supplementary section 2 **(Figure S4, Supporting Information)**. In brief, we placed the sample in between the feed and permeate chambers **(Figure 3a)**. While maintaining 100% $H_2O$ ($D_2O$) humidity, the $H_2$ ($D_2$) gas is directed towards the feed chamber where it converts to hydrons on the platinized carbon cloth ($H_2 \rightarrow 2H^+ + 2e^-$). We applied a voltage of ~1 V across the vermiculite device, which forced the protons/deuterons to pass through the vermiculite before recombining on the permeate side as $H_2$ ($D_2$) ($2H^+ + 2e^- \rightarrow H_2$). We observed that the mass spectrometer signal is absent when no voltage is applied across the devices **(Figure S5, Supporting Information)**. With applied voltages, there is a linear increase in signal, which ensures that our samples are leak-free.

With a reference sample, we observed that the concentration of the gas collected by our mass spectrometer is correlated with the measured electrical current (supplementary section 1, Supporting Information). The reference samples were analyzed with the mass spectrometer, and found that the ratio of $D_2$ to $H_2$, is ~ 0.93, which agrees well with our electrical measurements and the reported literature values[17]. In contrast, the vermiculite devices gave $D_2/H_2$ selectivity of 4.65 to 1.33 **(Figure 3b-c)**, very similar to electrical measurements. For checking repeatability and stability, each cycle was measured at least two times and found to be stable **(Figure S6, Supporting Information)**. This observation unambiguously confirms that $D_2$ is transporting faster than $H_2$ through vdW gaps of vermiculite laminates. Since the selectivity obtained from both the electrical measurements (Arrhenius plot) and the mass spectrometer is similar, our assumption of similar pre-exponential factors for both $D_2$ and $H_2$ in the Arrhenius relation (Equation (1)) holds true.

Having understood the transport of $H_2$ or $D_2$ across our membranes, we now performed the transport with an isotopic mixture. We exposed the samples to a 1:1 mixture of $H_2$: $D_2$ on the feed side (Inset in Figure 3c). We also kept a 1:1 mixture of $H^+$ and $D^+$ containing electrolytes ($H_2O$: $D_2O$ or HCl: DCl) on the same side to keep the Nafion hydrated (deuterated). A thin layer of gold was coated on the permeate side and was used as a cathode to complete the electrical configuration. This configuration is similar to previous reports[24]. A voltage of 1 V was applied across the sample for 12 hours before analyzing the collected gas signal with the mass spectrometer. As expected, in such cases, a signal at mass 3, corresponding to HD was seen and therefore accounted for in the selectivity calculations as per the following equation:

$$Selectivity\ \frac{D_2}{H_2} = \frac{(D_2 + \frac{1}{2}HD)}{(H_2 + \frac{1}{2}HD)} \qquad \text{Equation (2)}$$

In the reference Nafion sample, when exposed to a 1:1 mixture, the species detected are HD, $H_2$ and $D_2$ in decreasing order, and a reduction in selectivity from 0.93 to 0.73 was observed **(Figure S7, Supporting Information)**.

In contrast to the Nafion sample, our vermiculite sample shows higher permeation rates for $D_2$ (**Figure 3c**). However, the obtained $D_2/H_2$ selectivity reduced to 1.55 – 1.20 (inset in Figure 3c) instead of 4.65 – 1.33 (inset in Figure 3b). The device configuration used for these measurements is shown in **Figure S8, Supporting Information**. Given that all these measurements were performed at room temperature, this selectivity is significant (>1), indicating that deuterium transports more than hydrogen.

We also performed the isotope mixture measurements with the device configuration that is used for individual $H_2$ and $D_2$ measurements and found that the conclusions remained the same. For materials with a similar channel size and no selectivity suppression from the hydron selective electrode, better selectivity can be expected.

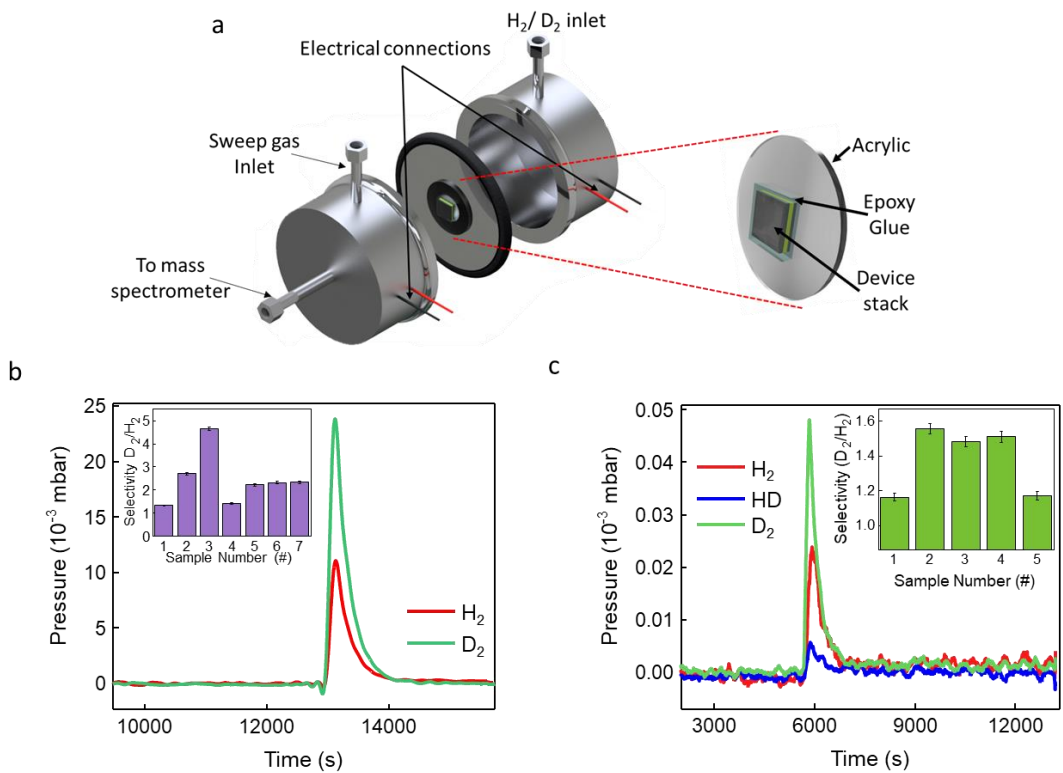

**Figure 3. Mass spectrometry measurements of vermiculite devices**. (a) Schematic diagram of the sample chamber used in mass spectrometry measurements. (b) $H_2$ and $D_2$ signals as detected by the mass spectrometer in separate measurements. The gas permeates through the sample under an applied voltage of 1 V and is stored for a duration of 3 hours to get a well-defined, detectable peak in the mass spectrometer. Under similar conditions, stronger deuterium signal is detected than hydrogen (original unsmoothed data in Figure S9, Supporting Information) (Inset: $D_2/H_2$ selectivity measured from seven samples, the error bars represent the standard deviation of the data). Both the peak values and the integral areas provide a similar $D_2/H_2$ selectivity. (c) The 1:1 isotope mixture measurement using a mass spectrometer. In contrast to separate measurements, a signal corresponding to HD is also observed. (Inset: The $D_2/H_2$ selectivity obtained in 1:1 isotope mixture measurement for 5 different samples, the error bars represent the standard deviation of the data).

The origin of this considerably large $D_2/H_2$ selectivity can be understood if we divide the whole hydron transport process into three steps. Initially, $H_2$ ($D_2$) converts to $H^+$ ($D^+$) on the platinized carbon cloth, where the presence of platinum helps in the dissociation of $H_2$ ($D_2$). The hydrons transport through the proton conducting Nafion layer in the second step. Here, the hydrons are not free but interact with the **Nafion's** sulfonate groups and $H_2O$ ($D_2O$). These two transport steps for the reference Nafion sample sufficiently explain the selectivity. In contrast, the vermiculite sample involves an additional transport path, where the hydrons pass through the vdW gaps. Of these three steps, the transport across the vdW gap is rate limiting, impacting selectivity.

In such smaller spaces, the wave nature of the isotope plays an important role. Deuteron has a smaller de Broglie wavelength than the proton, making it easier to pass through the confined 2.1 Å spaces than the proton. In agreement, we experimentally detected a higher amount of $D_2$ in the mass spectrometer measurements.

The experimentally observed selectivity ($\alpha$) values allowed us to calculate the confinement size encountered by the transporting isotopes theoretically. By using the following equation as described by Beenakker et al.[4]

$$\alpha = \exp\left(\frac{2\gamma_0^2\ \hbar^2\ \Delta m}{(d-\sigma)^2\ k_B T m_1\ m_2}\right) \qquad \text{Equation (3)}$$

where $\gamma_0$ is the zero of Bessel's function, $m_1, m_2$ are the masses of isotopes, $\Delta m$ is the mass difference of isotopes, $d$ is the confinement size, and $\sigma$ is the molecular hard core. Based on the selectivity in our measurements, we calculated the confinement sizes, which turned out to be 0.8 – 1.8 Å, surprisingly matching the vdW gaps obtained from the X-ray diffraction experiments. This also validates our hypothesis of kinetic quantum sieving effects in deuterium intercalated vermiculite laminates.

A recent study reports hydrogen isotope separation with monolayer lithium mica vermiculite membranes[31]. Due to the small interlayer spacing of lithium vermiculite (2θ ≈ 8.9°, interlayer spacing = 9.9 Å) with an effective vdW gap of 0.3 Å, no species can be transported. Instead, an $H_2/D_2$ selectivity of 6.7-9.8 was reported, and its origin has not been discussed in detail. We believe the transport is Knudsen-like, probably due to nano-sized defects in the membrane. In contrast, the transport across our vermiculite membranes happens through interlayer spacing. Our experiments consistently showed that the transport mechanism could be controlled by controlling the interlayer spacing and confinement, i.e., deuterium transport more than the proton, leading to a $D_2/H_2 >1$.

To gain further insights into the quantum sieving effects in 2D laminates in general, we investigated another laminate fabricated from graphene oxide (supplementary section 3, Supporting Information), which is well reported in the literature to have tunable interlayer spacing[32]. Like vermiculite, we found that graphene oxide interlayer spacing also reduces from 8.9 Å to 7.9 Å upon intercalation with deuterons **(Figure 4a)**. Nevertheless, in this case, the actual vdW gap available for the transport, calculated by subtracting the electron cloud (3.4 Å), is ~4 Å, much larger than that of vermiculite. In this case, the mass spectrometer measurements provided a $D_2/H_2$ selectivity of 1.17 **(Figure 4b)**, close to the theoretically estimated value of 1.06 as calculated using Equation (3). The exponential decrease in selectivity further strengthens the importance of confinement size in realizing the quantum sieving effect.

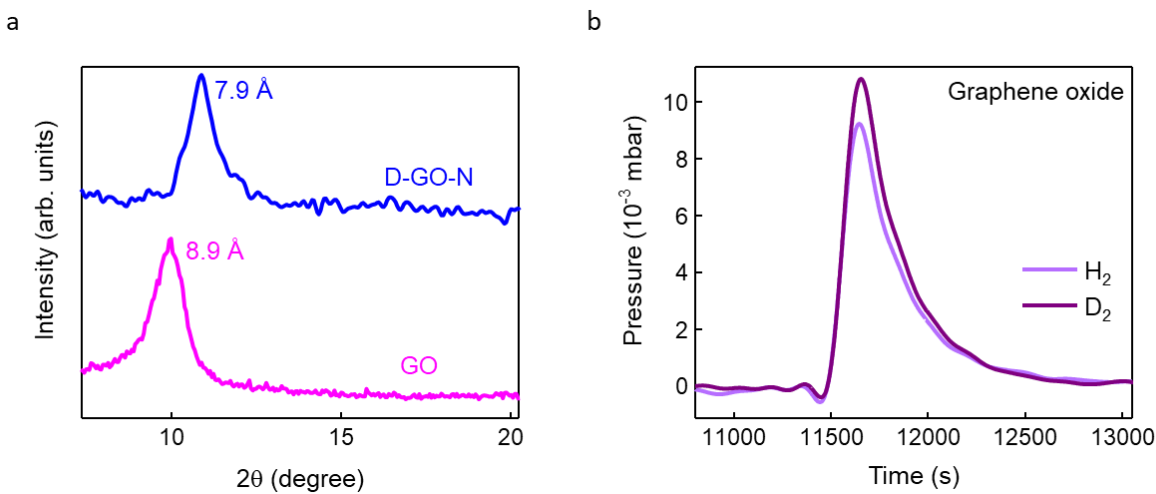

**Figure 4. Graphene oxide for hydrogen isotope separation.** (a) X-ray diffraction patterns of as-prepared graphene oxide laminate (magenta) and after intercalation with deuterium (blue). (b) Mass spectrometry measurements of proton and deuteron transport through a graphene oxide device (original unsmoothed data in Figure S10, Supporting Information).

3. Conclusions

We successfully demonstrated that deuterium intercalated vermiculite exhibits kinetic quantum sieving effect at room temperature leading to efficient extraction of deuterium. The cost-effective and scalable nature of vermiculite fabrication makes this an excellent choice for industrial applications. As a proof of the exponential nature of quantum sieving, we investigated graphene oxide, with a van der Waals gap of 4.0 Å and found that, when compared to 2.1 Å gapped vermiculite, the quantum sieving efficiency decreases exponentially. Thus, for further enhancing the quantum sieving effects, transport gaps much smaller than 2 Å are required. Laminates of two-dimensional materials offer tremendous opportunity to be explored further. Our work provides significant insights into the deuterium enrichment process, which will make nuclear energy production sustainable and economical.

**Supporting Information**

All data that is required to understand the conclusions in the paper is presented in the main and supporting information. Additional data related to this paper are available from the corresponding author upon request.

**Data Availability Statement**

The data that support the findings of this study are available from the corresponding author upon reasonable request.

**Acknowledgments**

This work was mainly funded by Science and Engineering Research Board (SERB), Government of India, through grant no. CRG/2023/004818. We also acknowledge the financial support from MHRD STARS with grant no. MoE-STARS/STARS-1/405 and also by DST-INAE with grant no. 2023/IN-TW/09. K.G. acknowledges the support of Kanchan and Harilal Doshi chair fund. The authors acknowledge the contribution from IITGN central instrumentation facility. L.H.Y. acknowledges the financial support from the National Science and Technology Council (NSTC), Taiwan under Grant No. NSTC 112-2923-E-011-003-MY.

**Author contributions**

K.G. conceived the idea and supervised the project. S.L. executed the sample preparation, characterization, electrical conductivity and mass spectrometry measurements. R.A. helped in initial laminate fabrication and mass spectrometer optimization. K.S. helped in XRD measurements. L.H.Y. helped in data analysis. S.L. and K.G. wrote the manuscript. All the authors discussed the results and commented on the manuscript.

**Declaration of interests**

The authors declare no competing interests.

**Supporting Information for**

**Room-Temperature Deuterium Separation in van der Waals Gap Engineered Vermiculite Quantum Sieves**

*Saini Lalita[1], Rathi Aparna[1], Kaushik Suvigya[1], Li-Hsien Yeh[3,4] and Kalon Gopinadhan[1,2*]*

[1]*Department of Physics, Indian Institute of Technology Gandhinagar, Gujarat 382355, India*
[2]*Department of Materials Engineering, Indian Institute of Technology Gandhinagar, Gujarat 382355, India*
[3]*Department of Chemical Engineering, National Taiwan University of Science and Technology, Taipei 10607, Taiwan*
[4]*Advanced Manufacturing Research Center, National Taiwan University of Science and Technology, Taipei 10607, Taiwan*

*(\*Correspondence: gopinadhan.kalon@iitgn.ac.in)*

**Methods:**

**Laminate fabrication**: For this study, the vermiculite crystals (Sigma Aldrich) are thermally expanded by using a two-step ion exchange method. In the first step, the crystals are heated in saturated NaCl (Finar) solution using a heat reflux. This step helps in exchanging the naturally present cations in vermiculite with $Na^+$ ions. In the second step, the crystals are washed with deionized (DI) water and then heated in 3 M LiCl (Finar) solution in heat reflux for 24 hours. With this, all the cations in the vermiculite crystals are exchanged with $Li^+$. Afterwards, the crystals are washed with DI water and dried in a microwave oven for 30 minutes. The dried crystals are weighed and dispersed in DI water (1 mg/ml), followed by bath sonication (Aztec) for 30 minutes to obtain a cloudy yellow suspension. To separate the unexfoliated particles, the dispersion is centrifuged (NEYA) at 4000 rpm for 10 minutes. Finally, a clear yellow suspension with exfoliated vermiculite nanosheets is obtained. This suspension is filtered through 0.22 µm pore size PVDF substrate (Sigma Aldrich) on a vacuum filtration assembly for fabricating the laminates.

After air drying, the laminates peel off from the support and are very flexible. For the intercalation of $K^+$ ($Al^{3+}$) ions in the interlayer spaces, the laminates are dipped in 1 M KCl ($AlCl_3$) for 24 hours. The salt ions provide hydrostatic stability to the vermiculite laminates and help in tuning the interlayer spaces. After the intercalation of salt ions, the laminates are washed with deionized water and dried with the help of an IR lamp (Murphy). The laminates are characterized using SEM and found to be free from any cracks or pin holes. The thickness of the laminates is estimated to be 3 µm using the cross-sectional SEM analysis.

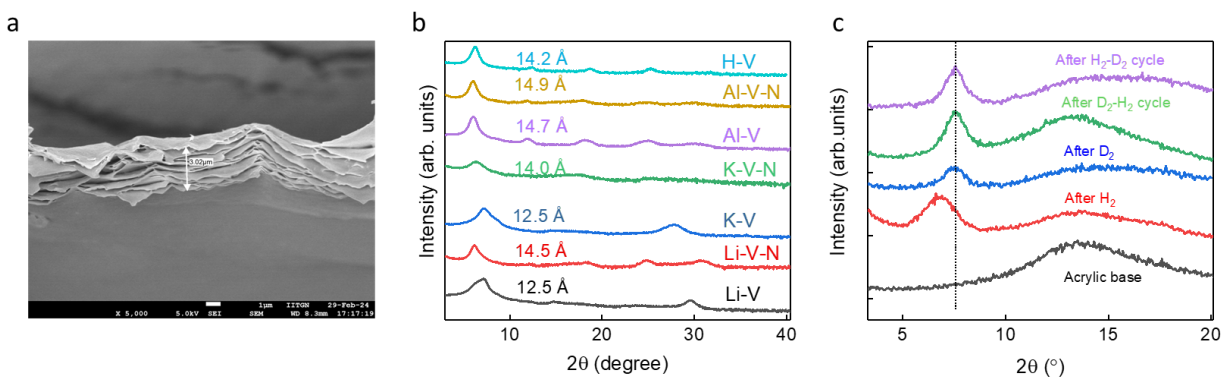

**Figure S1.** Characterizations. (a) Cross-sectional SEM image of a vermiculite laminate. (b) X-ray diffraction (XRD) pattern of vermiculite laminates intercalated with different cations and when coated with Nafion (-N). For the ion exchange, the fabricated vermiculite laminates were dipped in an aqueous solution (1 M) of various salt solutions for 24 hours. Afterwards, the laminates were washed repeatedly in deionized water to remove excess salt present on the surface, followed by drying under an IR lamp. To find the interlayer spacing of different ion-exchanged vermiculite, XRD (Rigaku SmartLab) was performed in glancing angle mode (glancing angle 1°). (c) The XRD pattern of the vermiculite after each cycle of hydrogen and deuterium exposure. Note that the interlayer spacing of vermiculite shrinks once deuterium is measured and does not change back

even when $H_2$ is exposed. In XRD, the acrylic base on which these devices are mounted gives a broad background running from 10 to 20 degrees.

**Supplementary section 1: Electrical conductivity measurements**

Electrical measurement setup:

For the vermiculite device fabrication, laminates are cut and pasted with epoxy glue (Stycast 2625) onto an acrylic support with a prefabricated hole of size 2 mm x 2 mm. The epoxy ensures that the only path for the ion diffusion is the interlayer space. Following this, the laminates are coated with Nafion solution (5%) and attached with electrodes made of platinized carbon cloth on both sides. For the electrical measurements, the device was kept in a custom-made sample stage connected to a blank flange KF40 with pins for electrical connections. Gold bonding wires (Tanaka, Japan) and silver paste were used to make contacts on both sides of the device.

The chamber used for electrical measurements was equipped with three valves, one for gas, liquid and vacuum each. After keeping the sample stage inside, the chamber was evacuated with the help of a dry scroll vacuum pump (Pfeiffer vacuum) for five minutes. Afterwards, the valves for gas and liquid were opened simultaneously for the flow of $H_2$ ($D_2$) and $H_2O$ ($D_2O$) inside the chamber. All valves were closed hereafter, and the samples remain in the hydrated (deuterated) gas environment. The humidity and temperature inside the chamber were continuously monitored with the help of a DHT22 sensor and a PT100 sensor, respectively. All the measurements have been performed at 100% RH unless stated otherwise.

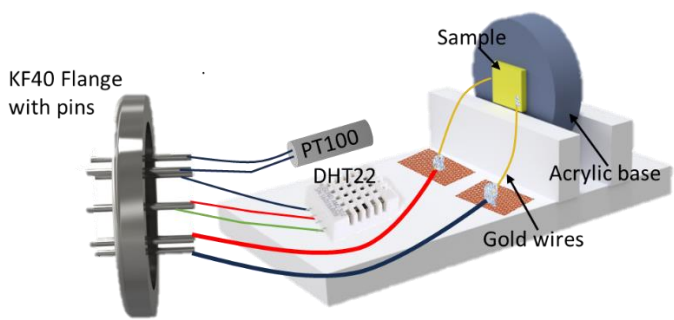

**Figure S2.** Schematic diagram of setup used in proton transport electrical measurements.

The current flowing through the samples was also monitored with the help of a Keithley 2614B source meter unit.

In the case of vermiculite devices, the first measurement involved the intercalation of $D^+$ inside the vermiculite interlayer spacings. As the intercalation proceeds, the current through the device was monitored continuously and was found to slowly increase before saturating around $1 - 5$ µA at an applied potential of 200 mV. At this point, the intercalation was considered to be complete, and the devices were used for further measurements.

For activation energy measurements, the temperature was varied with the help of a heater/ice bath and IV characteristics were measured. The temperature range was fixed between 20°C to 40°C for all the measurements to ensure a room temperature/accessible temperature operation.

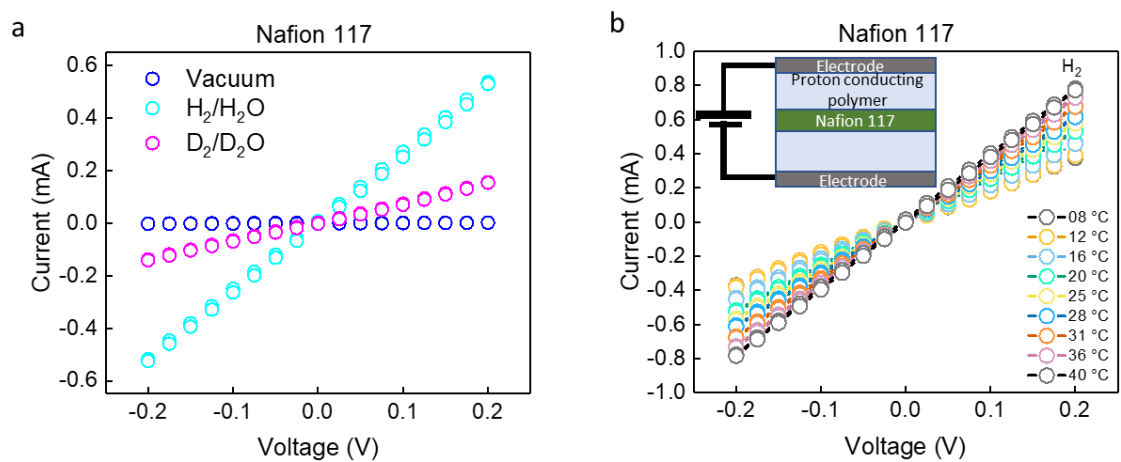

**Figure S3.** Control experiment with Nafion 117. (a) The current voltage characteristics of a Nafion 117 in vacuum and $H_2/H_2O$ and $D_2/D_2O$. The two-order difference in the current level of vacuum and hydrons shows that the measured current is solely related to the proton/deuteron current. (b) The current-voltage characteristics of a Nafion 117 at different temperatures in $H_2/H_2O$ environment. (Inset: Schematic diagram of device used for measurement).

Estimation of partial pressure of reference sample from the electrical measurements:

For estimating the expected pressure that we should get in mass spectrometer in the proton transport through our reference Nafion devices, we first measure the electrical current in the electrical measurement setup described above. For each hydrogen molecule transported through the device, we get 2 electrons contributing to the electrical current. Thus, we can estimate the number of hydrogen molecules transporting successfully as:

$$N = \frac{i}{2e}$$

where $i$ is the measured current and $e$ is electronic charge. Factor 2 in denominator accounts for the recombination of 2 protons and 2 electrons as one hydrogen molecule.

Now, we estimate the pressure using the ideal gas equation $pV = Nk_BT$, where $p$ is the pressure, $V$ is volume of the permeation chamber, $k_B$ and $T$ are the Boltzmann constant and the temperature, respectively. Considering the value of current through Nafion sample as 1.01 mA for 1 V, at room temperature (300 K), we calculate the estimated pressure as ~4.3 x $10^{-3}$ mbar. The calculated value agrees well with the measured values in mass spectrometer as described in the next section.

**Supplementary section 2: Mass spectrometer measurements**

Gas Transport measurement setup:

Before measuring the hydron transport in mass spectrometer, we first calibrated the mass spectrometer using a known feed mixture. For hydrogen isotope measurements, the m (mass)/z (valence) values of interest are 1 ($H^+$), 2 ($D^+$, $H_2^+$), 3 ($HD^+$) and 4 ($D_2^+$). We cannot use a 1:1 mixture of $H_2$ and $D_2$ for calibration because of the overlap of signals $D^+$ and $H_2^+$ at m/z of 2. We avoided this by using 1:1 mixture of $H_2O$ and $D_2O$, where the primary peaks are $H_2O^+$ (18) and $D_2O^+$ (20). The electron energy of the multiplier was optimized at 3 eV to enhance the detected signal. We observed primary peaks at m/z values of 18 and 20, corresponding to $H_2O^+$ and $D_2O^+$, respectively. Secondary peaks appeared at m/z values of 1 and 2, corresponding to $H^+$ and $D^+$ ions. By adjusting the electron energy window of the detector, we successfully obtained a 1:1 ratio for both the primary and secondary peaks of $H_2O$ and $D_2O$, following which we proceeded with the measurement of devices.

The devices used in our mass spectrometer measurements were made in a similar way to that of electrical measurements. After checking the conductivity, the devices were clamped in between the feed and permeate chamber of our custom-made assembly for such measurements (Figure. S4). For this purpose, we used epoxy glue to paste the device onto the O-ring, separating the two chambers. All devices were tested for leakage in dry as well as humid states to ensure the absence of non-selective pin holes. The assembly was custom-made to provide electrical connections inside the chambers so that electric field-driven hydron transport can be studied. After pasting on the O-ring, electrical connections were made with the help of a gold wire and silver paste before carefully clamping the O-ring in between the feed and permeate chamber. The electrical conductance was measured before and after assembling the cell to ensure the connections remained undisturbed. We also kept 1 ml of $H_2O$ ($D_2O$) inside both chambers to properly wet the Nafion so that the transporting particles face no extra resistance.

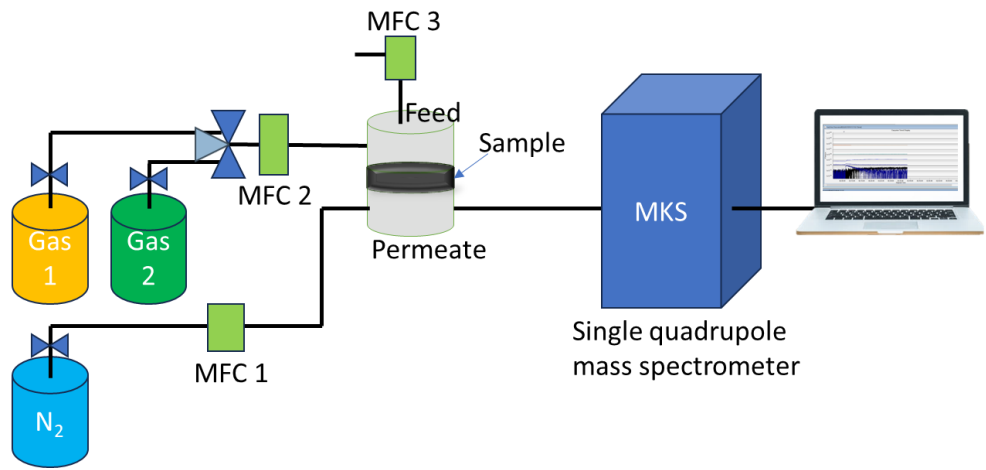

**Figure S4**. Schematic diagram of the assembly used for gas transport measurements.

After assembling the cell, the feed and permeate chambers were flushed with $N_2$ for an hour to remove any air present in the chambers, following which a 100% $H_2$ ($D_2$) was flowed at a flow rate of 4 ml/min into the feed chamber. The feed chamber was maintained at 1 bar pressure by flowing the gas out of the feed chamber at a similar flow rate. The permeate chamber was also maintained at the atmospheric pressure by flowing sweep gas ($N_2$) at a flow rate of 20 ml/min. The sweep gas carries the permeated particles to the mass spectrometer (MKS Cirrus 3-XD) while maintaining the permeate chamber at atmospheric pressure. The flow rates were controlled by using mass flow controllers (Aalborg DPC).

During this process, the current level was constantly monitored through the device until it reached a saturation value. At this point, the Nafion becomes completely saturated with the $H_2O$ ($D_2O$), and we apply a voltage across our device to facilitate the electric field-driven hydron transport. It should be noted that no signals for the isotopes were detected in the absence of voltage even after the pressure was increased above the atmospheric pressure which ensures that the driving force behind the transport is indeed the electric field and not the pressure difference. It also cross-checks that the molecular hydrogen and deuterium do not permeate though Nafion. The transport happens in the form of proton and deuteron which recombine as molecular hydrogen and deuterium before eventually be analyzed by the mass spectrometer. Therefore, the detected signals are a direct result of hydrons being permeated. No signal was detected in the absence of voltage which validates that our devices are leakproof (Figure S5).

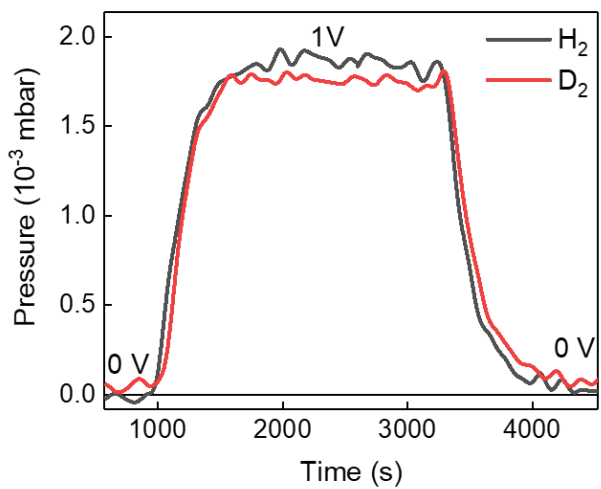

**Figure S5.** Detected signals of $H_2$ and $D_2$ through the Nafion device in separate measurements. No $H_2$ and $D_2$ signal can be detected in the absence of an applied electric field.

In between the cycles of $H_2$ and $D_2$, the device was flushed with $N_2$ on both sides until the background conductivity was achieved. As a control, devices made of commercial Nafion-117 membranes were measured to ensure the proper working of the setup, for which a $D_2/H_2$ selectivity of 0.93 was achieved (Figure S5). The reports in literature about hydrogen isotope selectivity ($H_2/D_2$) of Nafion range from 1-3[1,2], and this value aligns well within the range.

A similar procedure was followed for the measurement through vermiculite devices. But in this case, as the signal strength was estimated to be weaker than that of Nafion, we stored the permeated gas into the permeation chamber for 3 hours before analyzing with mass spectrometer. For each device, 2 to 3 cycles of measurement were done where $D_2$-$H_2$ is described as one cycle. For some devices, we detected a signal after the 2nd cycle without any applied voltage, which implied a leaky sample and were not measured further. For checking the stability, we stored the samples for more than 90 days and found that the samples remain stable.

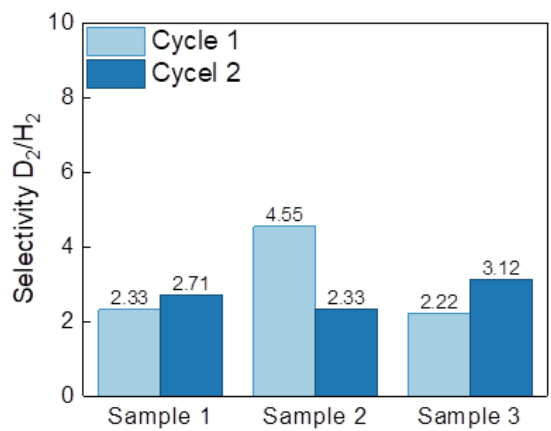

**Figure S6.** The selectivity measurements of two cycles from the vermiculite devices in mass spectroscopy measurements.

For finding the selectivity in the case of a mixture, a 1:1 mixture of $H_2$ and $D_2$ gas was used as the feed source. For keeping the Nafion properly hydrated (deuterated), a 1:1 mixture of $H_2O$: $D_2O$ or 1 mM HCl: 1mM DCl was used.

With an applied voltage, the mass spectrometer signal was observed in real time to find the fractions of permeated species. As expected, in such cases a signal at mass 3, corresponding to HD was detected, which was accounted in the selectivity calculations as per the following equation:

$$Selectivity\ \frac{D_2}{H_2} = \frac{(D_2 + \frac{1}{2}HD)}{(H_2 + \frac{1}{2}HD)}$$

In this way, the obtained selectivity was found to be 0.73, which is slightly lower than that calculated from separate measurements.

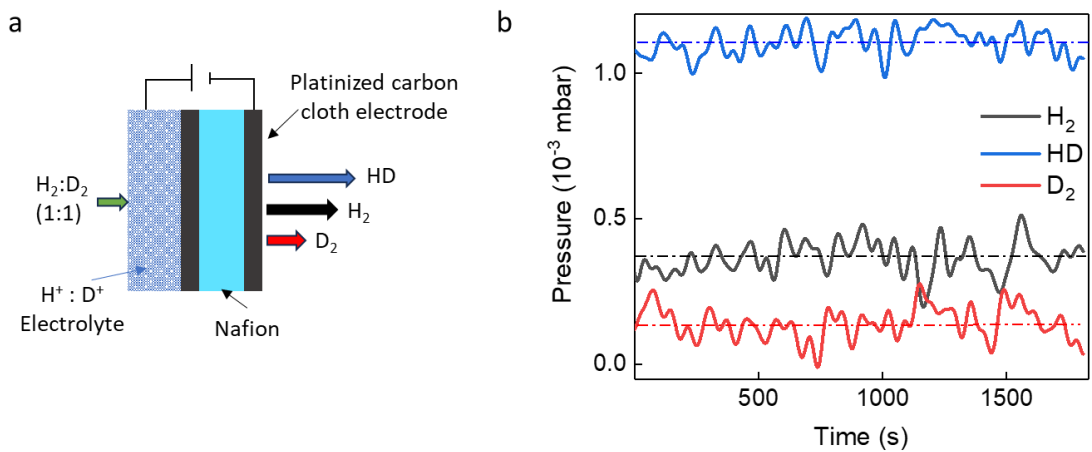

**Figure S7.** (a) Schematic diagram of the device configuration used for the hydron transport experiment in the case of mixture separation. The feed side was exposed to 1:1 mixture of $H_2$ and $D_2$, and 1 mM HCl + DCl electrolyte was used to wet the Nafion. When voltage was applied, the hydrons were transported to the permeate side. (b) Real time signal detection of $H_2$ (mass 2), $D_2$ (mass 4) and HD (mass 3) across the Nafion device with 1:1 mixture on the feed side and an applied voltage of 1 V. Dotted lines show the mean values of the signal and are used for the selectivity calculations.

Similar to the mixed electrolyte measurements of reference (Nafion) sample, in vermiculite also, a decrease in selectivity from 2 to 1.55-1.2 was observed.

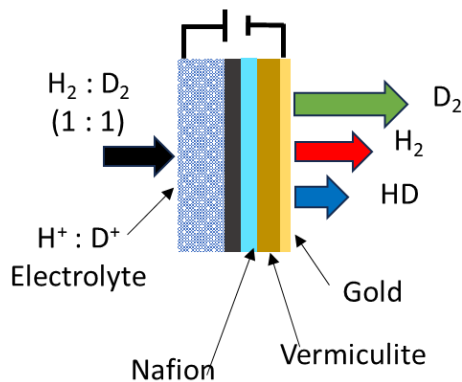

**Figure S8.** Schematic diagram of the device configuration used in isotope mixture measurement of vermiculite devices- The 1:1 isotope mixture of $H_2$ and $D_2$ is used as the feed source for isotopes and 1 mM HCl + 1mM DCl for proper wetting of Nafion. A gold electrode works as a cathode to complete the configuration

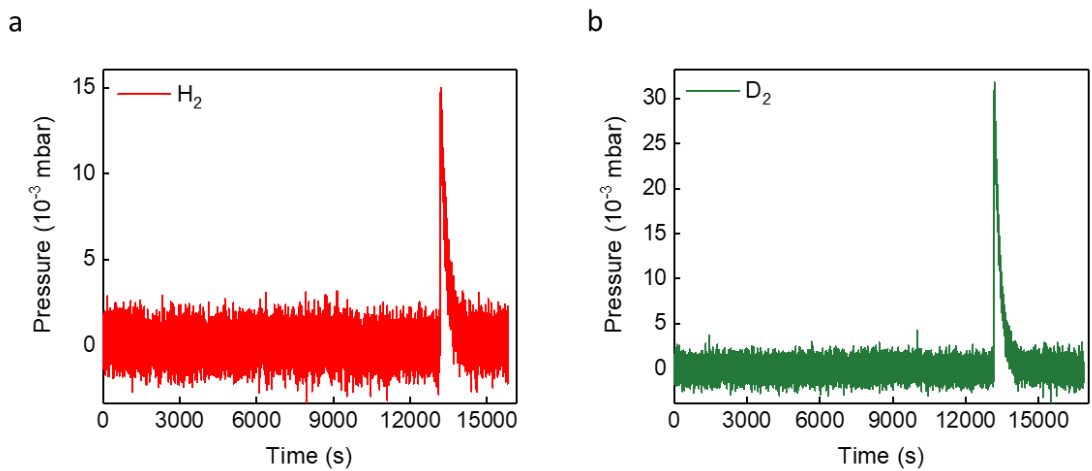

**Figure S9.** As-detected (raw, unsmoothed) signals of $H_2$ and $D_2$ through the vermiculite device during mass spectrometry measurements. The transported gas was collected for 3 hours in the permeation chamber before the analysis.

**Supplementary section 3: Graphene Oxide fabrication and measurements**

Fabrication of graphene oxide devices:

Graphene oxide (GO) was synthesized using modified Hummer's method. Briefly, graphite flakes (0.5 g) were added in concentrated $H_2SO_4$ (12 ml) under ice bath and continuously stirred at 1000 rpm. After five minutes, $KMnO_4$ (1.5 g) was added slowly, and a color change from black to green was observed. The ice bath was removed after 30 minutes allowing the reaction to continue at room temperature. After 6 hours, water was added drop-by-drop (50 ml) to stop the reaction. For removing unreacted $KMnO_4$, Hydrogen peroxide 35% (5 ml) was added to the mixture upon which a brown-yellow slurry of graphite oxide was obtained. The slurry was then washed with 10% HCl and deionized water until a pH of 6-7 was achieved. The graphite oxide slurry was then freeze dried, weighed and dispersed in water (1mg/ml) before sonication for one hour. After sonication, centrifugation was done to collect the dark brown supernatant containing graphene oxide flakes. 10 ml of supernatant was filtered on a PVDF substrate for making graphene oxide laminates. Upon drying, the laminates peel off from the substrate and are free standing, but unstable in water.

Next, a 1:1 solution of $Zn(NO_3)_2 \cdot 6H_2O$ (1 mM in methanol) and 2-methylimidazole (8 mM in methanol) was prepared and the GO laminates were dipped in it for 12 hours. In this step, ZIF-8 crystals are synthesized inside the interlayer spacing and increase the electrostatic interaction between the GO interlayers, which provide them stability inside the water. The stabilized laminates were washed with methanol twice and dried before using further for characterizations. For GO device fabrication and measurements, steps similar to that of vermiculite devices were followed.

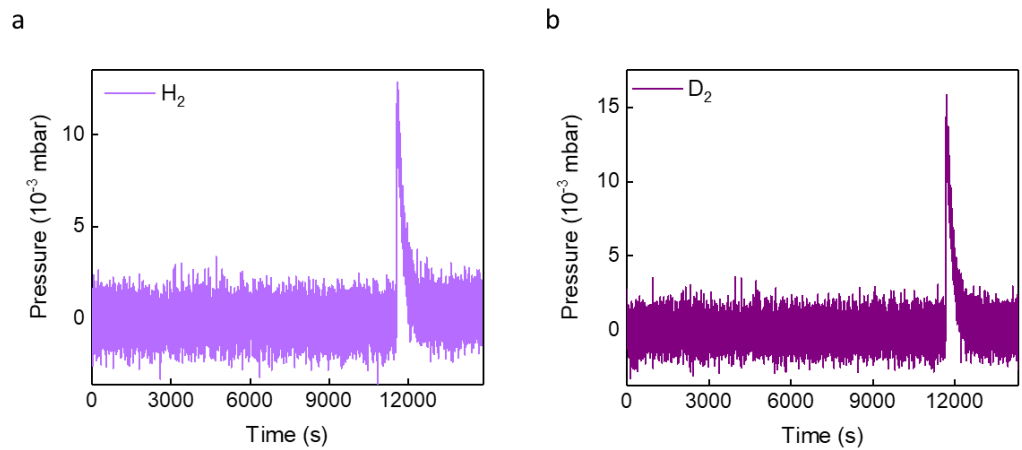

**Figure S10.** As-detected (raw, unsmoothed) signals of $H_2$ and $D_2$ through the graphene oxide device during mass spectrometry measurements. The transported gas was collected for 3 hours in the permeation chamber before the analysis.

**References**

1. Ba, J. *et al.* Inverse kinetic isotope effect of proton and deuteron permeation through pyridinic N-doped graphene. *Chemical Engineering Journal* **479**, 147423 (2023)

2. Lozada-Hidalgo, M. *et al.* Sieving hydrogen isotopes through two-dimensional crystals. *Science* **351**, 68–70 (2016).